\documentclass[preprint2]{aastex}

\usepackage{epsfig}

\newcommand{\II}{\protect\small II \normalsize $\!\!$}

\newcommand{\HII}{\mbox{\rm H\II}}
\newcommand{\Htwo}{${\rm H_2}$}
\newcommand{\I}{\protect\small I \normalsize $\!\!$}
\newcommand{\HI}{\mbox{\rm H\,\I}}

\newcommand{\inthms}[3]{$#1^{\rm h}#2^{\rm m}#3^{\rm s}$}

\newcommand{\intdms}[3]{$#1^{\circ}#2'#3''$}

\begin{document}

\title{Gas, Dust, and Young Stars \\
in the Outer Disk of M\,31}



\author{Jean--Charles~Cuillandre}
\affil{Canada-France-Hawaii Telescope, PO Box 1597, Kamuela HI 96743}

\author{James~Lequeux}
\affil{Observatoire de Paris, 61 Avenue de l'Observatoire, 75014 Paris, France}

\author{Ronald~J.~Allen}
\affil{Space Telescope Science Institute, 3700 San Martin Drive, Baltimore,
MD 21218}

\author{Yannick~Mellier and Emmanuel~Bertin}
\affil{Institut d'Astrophysique de Paris, 98 bis, bvd Arago,
75014 Paris, France; and \\
Observatoire de Paris, 61 Avenue de l'Observatoire, 75014 Paris, France}

\begin{abstract}

Using the Canada-France-Hawaii Telescope we have obtained deep high-resolution
CCD images in V and I of a $28' \times 28'$ field in the outer disk of M\,31 at
$\approx 116'$ from the center along the major axis to the south--west, and
covering a range of projected galactocentric distance from about 23 to 33 kpc.
The field was chosen to correspond with extended \HI\ features recorded near the
\HI\ edge of the galaxy.

The many tens of thousands of objects detected in this large field have been
classified using an automatic algorithm which distinguishes unresolved from
resolved structures and provides photometry on them.  For the most part the
unresolved objects are stars in M\,31.  The V--I colors of these stars are
highly correlated with the column density of \HI\ in the field.  Assuming a
Galactic extinction law, this yields a minimum extinction/atomic-gas ratio about
1/3 of that in the Solar neighbourhood.  The ISM in this outer disk of M\,31
therefore contains substantial amounts of dust.

We have identified a population of B stars in the field whose distribution is
also well correlated with the extended \HI\ distribution.  Evidently, star
formation is both ongoing and wide spread in the outer disk of M\,31.  According
to the current view of the star formation process, molecular gas is therefore
also expected to be present.

The objects classified as ``resolved'' turn out to be a mix of background
galaxies and overlapping images of foreground stars in M\,31.  The counts and
colors of the slightly-resolved objects in these ground-based CCD images
therefore cannot be used for a reliable determination of the total extinction
and reddening by the interstellar medium in M\,31.  However, the larger
background galaxies are easily recognizable, and their surface density above a
specific magnitude limit is anti--correlated with the \HI\ column density,
confirming that a relatively large amount of extinction is closely associated
with the \HI\ gas.

\end{abstract}

\keywords{galaxies: individual: M\,31 --- galaxies: ISM --- 
galaxies: star formation --- ISM: dust, extinction: stellar populations}

\section{Introduction }

How ``primordial'' is the interstellar gas in the far outer parts of galaxies?
Does this gas show any signs of metal enrichment?  How much dust is there in the
outskirts of galaxies?  Is there any evidence for substantial quantities of dark
matter in the form of \Htwo\ in these regions?  The answers to these questions
are relevant to our understanding of galaxy formation and evolution and the
chemical enrichment of the interstellar medium over time.  However, the low
densities and faint radiation fields in the outskirts of galaxies make it
difficult to detect dust in the interstellar medium (ISM) there.

There are strong incentives to attempt a detection and quantification of the
amount of dust present in the outskirts of galaxy disks.  Large amounts of dust
might indicate indirectly the presence of large quantities of otherwise
undetectable molecular hydrogen.  Cold molecular hydrogen has been suggested as
a possible component of dark matter in the disks and even the halos of spiral
galaxies.  It is not clear that this \Htwo\ would have to be associated with
\HI, so that dust uncorrelated with \HI\ might indicate the presence of \Htwo.
On the other hand, when dust is observed to be correlated with \HI, one can hope
to determine the dust-to-gas ratio and, indirectly, the metallicity of the gas
(e.g.~\citet{bouc85}).  This may offer the best way to determine the metallicity
in the outer regions of galactic disks, where there are generally no bright
\HII\ regions.

\citet{hod74} used visual counts of galaxies to map the extinction in the SMC,
comparing this to the \HI\ in order to obtain the gas/dust ratio.  \citet{mac75}
did the same work in a more detailed way and found regions of high absorption at
the edges of the SMC.  Using these results, \citet{leq94} suggested that a large
amount of molecular hydrogen exists in the outer parts of the SMC.  The colors
of background galaxies were used by \citet{zar94}, and later by \citet{leq95} to
search for dust in the halos of galaxies.  The first convincing results on disk
galaxies were obtained by \citet{gon98} using bright background galaxies
identified by their appearance and colors on HST images of a nearby foreground
galaxy.  \citet{gon98} obtained the reddening from galaxy colors and the
extinction from galaxy counts, using a method of comparison with
\textit{synthetic fields} in order to calibrate the effects of confusion on the
search and classification algorithms.  The high angular resolution of HST was
crucial to the success of this method.  However, careful visual examination of
all candidate galaxies was also required, and this method is therefore quite
time-consuming.

In this paper, we discuss the reddening and extinction in a field far out in the
disk of M\,31, based on observations of M\,31 stars and of background galaxies.
A by--product of this observation is the photometry of a large number of stars,
from which high--quality color--magnitude (C--M) diagrams can be constructed, as
well as maps of different categories of stars.  Section 2 describes
briefly the observations and the calibrations.  In \S 3 we
describe the analysis of the data, and in \S 5.2 we determine a
lower limit to the extinction from the C--M diagrams of the stars.  Section
6 discusses the distribution on the sky of relatively bright
background galaxies and their extinction and reddening.  In \S
7 we show the locations of young blue stars found in the field
and discuss the relation of these young stars to the \HI.  Our conclusions are
summarized in \S 8.

\section{The Data }

\subsection{Observations }

We observed a field of $28' \times 28'$ with the large-format UH8K CCD mosaic at
the prime focus of the Canada-France-Hawaii Telescope.  The field is centered at
\inthms{00}{36}{66}, \intdms{+39}{48}{50}, J2000.0 and located $\approx 116'$
from the center of M\,31 along the major axis to the south--west\footnote{We use
the conventional distance of 690 kpc for M\,31, in which case $1' = 200.7$ pc
along the major axis of the galaxy.  A very recent estimate puts M\,31 at a
distance of 890 kpc \citep{fea99}, in which case $1' = 258.9$ pc.}, covering a
range of projected galactocentric distance from approximately 23 to 33 kpc.
This region is at the extreme limit of the visible disk of the galaxy on the SW
side, with $\mu_{\rm B} \gtrsim 26.5$ mag arcsec$^{-2}$, and extends from 4 to
5.7 times the disk scale length in the B band \citep{inn82,wal88}.  It also
includes the outer boundary of the \HI\ disk as observed by \citet{new77}.
Figure 1 is a sketch of the location of our full UH8K CCD field
of view on the image of M\,31.

The observations were made in October 1995 with a mosaic camera consisting of 8
CCDs, each $2048 \times 4096$ pixels, arranged into a $4 \times 2$ array
totalling $8192 \times 8192$ pixels.  The pixel size corresponds to $0.206''$ on
the sky.  Unfortunately, one of the eight CCDs (the chip in the north--west
corner) was not of the same quality as the others, and we have ignored this part
of the field.  The final UH8K field is shown in Figure 2 with
a grid overlaid to show the disposition of the eight CCD chips.

Our observations consisted of a series of 20-minute exposures collected over 4
nights at the telescope, 16 in I and 17 in V.  Of these, 10 in I and 11 in V
were of good photometric and seeing quality, giving total exposure times of 200
and 220 minutes, respectively.  The filter responses are uniform over the whole
field and are close to the Cousins system.  The 8 thick CCDs come from a similar
batch and have the same response, which is uniform across the chip.  The
individual exposures were slightly shifted in position with respect to each
other in order to fill in the gaps between the individual chips and to permit
the identification of bad pixels and cosmic rays.  The separate exposures were
then regridded onto a common $0.206''$ grid, recentered, and averaged after
clipping aberrant pixels.  This procedure effectively eliminated bad pixels and
pixels affected by the impacts of cosmic rays, while retaining the correct
photometry.  The point spread function (PSF) on the combined exposures is
$\approx 0.70''$ FWHM in both bands.

At the time these observations were made (October 1995) the capabilities
available in the standard data reduction systems such as IRAF and IDL were
inadequate to cope either with the volume or with the mosaic format of the UH8K
CCD camera data.  Furthermore, the UH8K camera is only one step in a larger
program of wide-field CCD imaging at the CFHT aimed at commissioning a camera
with $18,000 \times 20,000$ pixels in 2002 (MegaCam, \citet{boul98}).  A new
image processing system (FLIPS, FITS Large Images Processing Software) was
therefore developed to deal efficiently with the new large mosaic data sets
\citep{cui01}.

\subsection{Calibration }

Each CCD was separately calibrated by observing standard stars in the SA\,113
field \citep{lan92}.  These stars span a reasonable range in color, so that
color equations can be computed.  The zero points and colors were obtained by
using a single setting of the mosaic camera on the field, providing different
SA\,113 stars on each CCD.  Flat--fielding was done with a library ``superflat''
resulting from the combination of all the available observations made with this
mosaic camera in the week preceding our M\,31 observing run.  However, some
small discrepancies remain (maximum 4\%) owing to instabilities in the CCDs and
to small errors in the color equation.  A final correction was made by adding a
small constant to each CCD in order to align the V--I colors of the pixels along
neighboring edges of each CCD.  The final internal standard error for each CCD
on the calibration stars is 0.05 mag in V and 0.04 mag in I, while the standard
deviations between CCDs are 0.02 and 0.04 mag, respectively.

As an example of the detailed image characteristics we show in Figure
3 a portion of the final averaged V image from CCD3.  This
image illustrates the high density of M\,31 stars, as well as showing several
bright galaxies.

\section{Image Analysis }

We have carried out an automatic classification of objects in the entire field
using a recent version of the algorithm SExtractor \citep{ber96}.  The task
PSFex was used to identify stellar objects on the image and to perform stellar
photometry.  ``Stars'' were defined as objects with ${\rm FWHM} < 1.1''$ (the
PSF is estimated to be $0.70''$) and ellipticity $< 1.2$.  ``Galaxies'' were
defined as objects with ${\rm FWHM} > 1.5''$ and any ellipticity; the task
MAGAUTO in SExtractor was used for photometry of the galaxies.  The result is a
catalog of stars and a catalog of candidate galaxies with V and I magnitudes.

In total, more than 170,000 objects were catalogued in two colors on 7 of the 8
UH8K CCD chips.  In Figure 5a we show the histograms of the
magnitudes of all objects catalogued on a representative subfield of area 100
arcmin$^2$ from CCD1 on the final image.  These histograms show that the counts
of detected objects rise smoothly up to the magnitude limits adopted, and that
our sample is therefore essentially complete for $V < 25.4$, $I < 24.4$.  In
Figure 5b we show the histogram of the widths (FWHM) of
objects in the same subfield, confirming a high degree of completeness in our
sample for FWHM $ > 0.7''$ as well as justifying our choice of the size limit
for stars (FWHM $ < 1.1"$).  Unfortunately, as we shall see in the next section,
the ``shoulder'' in this histogram from roughly $1.4''$ to $2.3''$ is coming
mostly from overlapping star images and not from faint background galaxies.

\section{Counts of Small Field Galaxies }

We constructed C--M diagrams for those objects classified as ``galaxies'' in
various parts of our M\,31 field, and compared the results to data obtained in
other programs with the same mosaic camera on ``blank'' fields located far from
any nearby galaxy.  We first describe the results on the blank field.

\subsection{Galaxies in a ``blank'' field }

The blank field we have used for comparison with M\,31 is in SA\,57 near the North
Galactic Pole \citep{maj94}.  It was observed for several hours both in V and I
for another project, using the same combination of instrumentation as we have
used for M\,31.  These data were also reduced, combined, and calibrated by one
of us (J.C.) with the same procedures we have used for M\,31.  A portion of the final
averaged image field in V is shown in Figure 4.

Figure 6 shows a C--M diagram for all blank-field galaxies in
CCD3 of the UH8K mosaic.  The numbers along the left side of the plot give the
number of galaxies in the relevant magnitude interval.  In Table
1 we collect some data for counts and colors of blank field
galaxies which will be useful later.  Above the completeness limit of $V <
24.75$, there are 2025 \textit{bona fide} galaxies larger than $1.5''$ in this
$7.03' \times 14.06'$ field.

\subsection{``Galaxies'' in the M\,31 field }

The comparison of our results on objects classified as galaxies in the M\,31
field with the blank field described above shows that our ``galaxies'' are
severely contaminated by confusion from overlapping images of stars in M\,31,
since there are significantly more ``galaxies'' per unit area in our M\,31 field
than there are in blank sky fields.  For instance, in the part of the C--M
diagram limited by $V < 24.75$ where incompleteness is not a serious problem,
the classification algorithm finds $\approx 6000$ candidate galaxies per CCD
frame in M\,31, whereas the same algorithm finds only 2025 \textit{bona fide}
galaxies larger than $1.5''$ in the same area of the blank field C--M diagram
(Figure 6).  Under these conditions, it is clearly hopeless to
try to use counts and colors of faint background galaxies in order to determine
reddening and extinction in our M\,31 field.

However, we shall see that the C--M diagrams for \textit{the stars} are very
clean (e.g.\ Figures 8a and b to be described below), and can be
used to determine at least a \textit{lower limit} to the total opacity through
the outer disk of M\,31.  Also, analogous to the method used by \citet{gon98},
bright background galaxies are large enough to be recognized unambiguously in
our field and are not too numerous to count visually, so that estimates of
variations in number density and colors of these bright galaxies can be
determined.

\section{C--M Diagrams of the Stars in M\,31 }

As a first comparison of the stellar color--magnitude (C--M) diagrams in
gas-rich and gas-poor parts of the field, we have defined two regions which are
representative of the lowest and highest \HI\ content.  Figure 7
shows these areas with respect to the \HI\ contours.  These regions have the
same area on the sky, 70 arcmin$^2$.

Figure 8b shows a diagram of V vs.\ V-I for the faint-\HI\ area.
The boxes drawn on the Figure show the approximate locations of B stars and Red
Giants (RG) expected in M\,31 for a distance modulus of 24.19 mag, corresponding
to our adopted distance of 690 kpc.  Figure 8a shows the
corresponding C--M diagram for the region of high \HI\ column density.  This
region includes the sky within the 4th brightest contour on the \HI\ data, i.e.\
average \HI\ columns in excess of $N(\HI) = 1.1 \times 10^{21}$ \HI\ atoms
cm$^{-2}$.  This figure differs in two important ways from the C--M diagram in
the low-\HI\ areas:  First, the bulk of the stars appear to be both slightly
\textit{redder} and slightly \textit{fainter}, indicating the presence of
obscuring dust; and second, \textit{a population of B stars has appeared},
indicating recent star formation.

We have made a rough estimate of the ``reddening law'' in this part of M31 by
determining visually the shift in V and V-I from Figure 8b to
Figure 8a.  The process is sketched in
Figure 9.  The approximate
result is E(V-I) $ \approx 0.17$, A$_{\rm V}$ $ \approx 0.35$, so that
E(V-I)/A$_{\rm V}$ $\approx 0.5$, which is very close to the (more precise)
value of 0.52 for dust in the Galaxy \citep{sav79}.

We note that our estimates of star counts in M\,31 are contaminated by
unresolved background galaxies.  However, this effect is small; for $V < 24.75$
there are $\approx 2000$ galaxies per CCD frame in the reference field we have
taken far from any foreground galaxies, as described in \S
4.1.  But in our M\,31 fields there are $\gtrsim 20,000$
stars per CCD frame at this magnitude level.  The contamination by unresolved
background galaxies in Figures 8a and 8b is
therefore negligible.

\subsection{Color variations of the stars }

In order to quantify the variations in the colors of the stars with location
over this part of M\,31, we split the entire image up into small square
sub-areas of size $340 \times 340$ pixels = $1.167' \times 1.167' = 1.36$
arcmin$^2$ which we call a ``superpixel'', large enough to contain a sufficient
number of stars for reasonable statistics (about 230 stars within the magnitude
and color limits adopted), but small enough to show variations on the arcminute
scale over the image.  For each superpixel we build the histogram of V--I
colors.  The rms dispersion $\sigma$ is measured, and the histogram clipped at
$2 \sigma$ in order to minimize the effect of the histogram wings which are
contaminated by objects with peculiar colors and affected by incompleteness at
the faint magnitudes.  We then calculate the average color $\langle V - I
\rangle$ for the remaining part of the histogram\footnote{Of course the colors
of individual stars depend on their magnitudes as well, which is clear from the
C--M diagram in e.g.\ Figure 8a.  However, we care only about the
variation in \textit{average} colors with position in the field.  Since the
slope of the C--M relation for Red Giants is relatively constant over the field
(and the same as that in many other fields both in M\,31 and outside of it), our
results on the average colors above our magnitude limit are not likely to be
affected.}.  Finally, to ensure a reasonable degree of completeness and to avoid
objects not likely to be M\,31 stars, we have limited our analysis of the
superpixel C--M diagrams to the range $V < 25.4$, $20.5 < I < 24.4$.  These
cutoffs were chosen to keep the sample complete over the whole mosaic, although
the most constraining chip is CCD2 in the upper left of the mosaic closest to
the main body of M\,31.  For the analysis of average colors of the old disk
stars we also want to avoid including blue stars, since their density appears to
vary over the image (cf.\ \S 7); for this analysis we
therefore limit the range of object colors to $0.5 < V-I < 4$.

In Figure 10 we show representative C--M diagrams for two
superpixels, one in an \HI-rich region, and the other in an \HI-poor region,
along with their accompanying histograms.  The $2 \sigma$ clip levels are shown
as vertical lines in the histograms, and the short horizontal arrows are $1
\sigma$ in length and anchored in a large dot showing the mean color.  Compared
to the histogram of the \HI-poor area, the histogram in the \HI-rich region has
about the same shape and location on the blue side, but is noticeably wider on
the red side, with two secondary peaks extending the distribution about 0.5 mag
further to the red.  For the present purposes we use only the shift in the mean
color, but we note that the shape of the histogram in the \HI-rich area suggests
a clumpy distribution of obscuration, on length scales smaller than the $\approx
200$ pc size of a superpixel, and with opacity values up to two or more times
the mean opacity over the superpixel.

\subsection{The correlation of stellar {\protect $\langle V - I \rangle$} \\ with N({\protect \HI})}

A map of the superpixel $\langle V - I \rangle$ for with $V < 25.4$, $20.5 < I <
24.4$ was constructed and then smoothed to $3 \times 5$ superpixels = $3.5'
\times 5.85'$, approximately the same angular resolution as the \HI\ map of
\citet{new77} ($3.6'\:  {\rm EW} \times 5.8' \:  {\rm NS}$ FWHM)\footnote{Note
that the map shown here in Figure 11 has been corrected for the
primary beam response of the radio telescope used to produce it, contrary to the
maps published in the original paper by \citet{new77}.}.  The two maps are shown
superimposed in Figure 11, where the contours represent N(\HI) and
the grey scale $\langle V - I \rangle$.  A good correspondence appears to exist
between the reddest superpixels and the highest column densities of \HI.  This
indicates not only that there is dust mixed in with the \HI\ in these outermost
parts of M\,31, but that the dust is of such a nature as to cause at least some
measurable additional reddening of the light from the stars embedded in it.  The
reddening deduced here is, of course, only a lower limit to the total reddening
through the disk of M\,31, since some fraction of the stars we are using will be
in front of the absorbing material.

In order to determine the amount of dust extinction present in the \HI\ cloud in
Figure 11 we must make some assumption about the reddening law
connecting the excess reddening E(V-I) of starlight with the extinction A$_{\rm
V}$.  For the Galaxy this is E(V-I)/A$_{\rm V}$ = 0.52 \citep{sav79}.  In
section 5 we have estimated this value to be $\approx 0.5$ for
our M\,31 field.  However, we prefer to use the more accurate Galactic value,
since our determination is only approximate.  This means that extinction e.g.\
at V moves the representative points in the V, V--I diagram not at constant V,
but along reddening lines of slope E(V-I)/A$_{\rm V}$ = 0.52 (see e.g.\ Figure
9).  This introduces a small correction since the
true E(V-I) is therefore only about 90\% of that obtained by considering the
change in $\langle V - I \rangle$ as a function of position at constant V.  We
have made this correction in what follows.

A further potential problem is the fact that V--I also depends on the
metallicity of the stars as well as on the characteristics of the intervening
dust.  We do not expect to find much change in metallicity over the small part
of the M\,31 disk appearing in our image, but the V--I color will become bluer
when one enters the halo.  It is of course not possible to exclude completely
the contribution of halo stars in each superpixel, but we can try to minimize
their effects on the color gradient.  We have therefore eliminated the 9 columns
of superpixels on the western edge of Figure 11 where more rapid
changes in the relative numbers of disk and halo stars are expected.  Over the
rest of the image the disk stars dominate and the contribution from halo stars
to the V--I colors of each superpixel ought to be minimal.

The final correlation between E(V-I) and N(\HI) obtained from our data is shown
in Figure 12.  Note that the points in this figure are not all
independent, hence the apparent non-gaussian distribution.  In order to compare
this result to the values found for the Galaxy we need to convert our E(V--I)
values to E(B--V).  Assuming again that the extinction law in this part of M\,31
is the same as that for solar neighbourhood of the Galaxy\footnote{This is a
reasonable assumption according to \citet{bouc85} who find similar laws in the
visible and the infrared for the Galaxy and the two Magellanic Clouds in spite
of large differences in metallicity.}, one has E(V-I$_{\rm Cousins}$) $ = 1.60
\times$ E(B-V).  Thus, including the 90\% correction factor explained above, the
correlation in Figure 12 implies:

\begin{equation}
\frac{\rm N(\HI)}{\rm E(B-V)} = 1.45 \times 10^{22}\:
{\rm atoms}\:{\rm cm}^{-2}\:{\rm mag}^{-1},
\end{equation}

\noindent which is 3 times the canonical Galactic value of $4.8 \times 10^{21}$
atoms cm$^{-2}$ mag$^{-1}$ \citep{boh78}.  We recall that the reddening we
measure is only a lower limit to the total reddening through the disk of M\,31,
since some fraction of the stars we are using will be in front of the absorbing
material, whereas the \HI\ column refers to the entire line of sight through the
galaxy.  The value for N(\HI)/E(B-V) we give above is therefore an overestimate,
and there will be more dust present on average than we have determined.

Our lower limit to E(B-V) shows that, even in these sparse outer parts of M\,31,
the \HI\ still contains a large amount of dust, indicative of a metallicity at
least 1/3 of the value in the solar neighborhood if the dust/gas ratio is
proportional to metallicity (e.g.  \citet{iss90}; \citet{sch93};
\citet{bouc85}).  An alternative to high dust/gas ratio or high metallicity
could be that the region contains not only atomic, but also molecular hydrogen,
with a distribution similar to that of \HI.  The dust we have detected would
then correspond to both components of the gas, whereas our estimate of the gas
column of course refers only to the atomic component.

\section{The Bright Background Galaxies }

Owing to confusion by overlapping star images in M\,31, small galaxies can not
reliably be identified for use as probes of obscuration in our image.  However,
there are quite a number of ``large'', bright galaxies which appear in the
image; they cover many pixels and are easily identified.  We have therefore
catalogued these objects by a detailed visual examination of each CCD chip and
measured their magnitudes and colors.  The identifications were independently
made by two of us (J.C.\ and J.L.)  and are considered extremely reliable.  The
number of these objects appearing in various magnitude intervals over the entire
field of all 7 CCD chips is as follows:  $V < 19.5$, 66 objects; $19.5 < V <
20.5$, 193; $20.5 < V < 21.5$, 148; and, $21.5 < V < 22.5$, 147.  The locations
of these galaxies are shown in Figure 13 along with the
contours of N(\HI).  Note the striking lack of bright galaxies just in the
region of the highest \HI\ contours in this figure.

\subsection{Counts of the bright galaxies }

The deficit of bright galaxies at high values of N(\HI) can be quantified by
counting the number of galaxies with $V < 22.5$ in Figure 13 at
various \HI\ contour intervals.  The results are shown in Figure
14a.  The data show a slight downward trend of the counts
with increasing N(\HI); a linear regression through the points gives:

\begin{equation}
N_G = 0.725 -1.6 \times \frac{N(\HI)}{10^{22}\:{\rm cm}^{-2}},
\end{equation}

\noindent in units of galaxies/superpixel, with a correlation coefficient 0.828.
This relation can be converted into a relation between extinction and N(\HI) as
follows:  The differential luminosity function of galaxies at a given magnitude
can be written as $N_G(m) = C \times 10^{\alpha m}$.  The ratio of the counts
$N_G(m)$ in one extincted region (e.g.\ with extinction A$_V$ mag) to those
$N_{G0}(m)$ in a reference region at the same detection limit is then
$N_G(m)/N_{G0}(m) = 10^{-\alpha A_V}$.  Now $\alpha = 0.38$ \citep{arn97} for
galaxy counts in V over a similar range of magnitudes ($19 < V < 24$).  Hence
from $N(\HI) = 0$ to $15 \times 10^{20},$ $N_G(m)/N_{G0}(m) = 0.67$ and so $A_V
= 0.46$ mag.  This yields $N(\HI)/A_V = 3.3 \times 10^{21},$ or $N(\HI)/E(B-V) =
1.0 \times 10^{22}$, to be compared with $4.8 \times 10^{21}$ for our Galaxy.
Thus the dust-to-atomic-gas ratio by this method is $\approx 0.5$ Galactic,
compared to the minimum value of $\gtrsim 0.33$ Galactic found from the stars.
These results are therefore consistent, although the errors on the counts of the
bright galaxies are of course much larger.

It is interesting to note that the decrease in large galaxy counts with
increasing $A_V$ which we have found in the outer parts of M\,31 is entirely
consistent with the results of \citet{gon98}, who applied a much more exhaustive
analysis to fields of nearby galaxies imaged with the WFPC2 camera on the Hubble
Space Telescope.  Our equation 2 above can be re-written as
$N_G/N_{G0} = 1 - 0.81 \times A_V$, while the simulations shown in Figure 13 of
\citet{gon98} can be parametrized by $N_G/N_{G0} = 1 - 0.77 \times A_V$ at least
out to an $A_V \approx 1$, after accounting for the small excess reddening of
$E(V-I) \approx 0.2$ which they determined from their data on NGC\,4536.
\citet{gon98} also found evidence for a contribution of high opacity clumps, at
least in spiral arms.

Could the deficit in the distribution of bright galaxies at high values of
N(\HI) shown in Figure 13 be merely a consequence of the known
\textit{clustering of galaxies} in the Universe?  While this can not obviously
be ruled out, we note that a two-fold conspiracy is required:  First, the
position of the ``void'' on the sky would have to correspond closely with the
distribution of N(\HI); and second, the decrease in galaxy counts shown in
Figure 14a would have to be just what would be predicted
from the distinctly different analysis of \citet{gon98} on an entirely different
galaxy.

\subsection{Colors of the bright galaxies}

Our data analysis also yields the colors of the bright galaxies.  We have
calculated histograms of the V--I colors for all bright galaxies in each
interval of N(\HI).  The histograms show some outliers, perhaps a consequence of
inaccurate photometry, so we have generally used the median colors instead of
the means.  If the means were used, we first truncated the histograms at
$2\sigma$.  The results are shown in Figure 14b, from which
we have to conclude that there is no evidence for reddening of the bright
background galaxies, whereas the reddening of the stars was 0.15 mag for the
same conditions.  This result is actually consistent with the behavior of galaxy
colors at faint magnitudes.  The effective magnitude limit will shift at higher
values of N(\HI) owing to the extinction.  For instance, at $N(\HI) = 1.5 \times
10^{21}$ cm$^{-2}$, $A_V \approx 0.4$, so the magnitude limit of $V < 22.5$ is
actually shifted to $V < 22.1$.  Of course it is precisely this shift which
causes the decrease in the counts shown in Figure 14a.  But
this shift also accompanied by a change of \textit{intrinsic} color of the
background galaxies.  From the data in the comparison field in Figure
6, and summarized in Table 1, we see that a shift
in the magnitude limit from 22.5 to 22.1 leads to an average \textit{blueing} of
galaxies by $\Delta\langle V-I \rangle = -0.1$, just enough to compensate for
the anticipated \textit{reddening} of $\Delta\langle V-I \rangle = 0.15$
observed for the stars with the same amount of extinction.  We conclude that the
absence of reddening with increasing N(\HI) for the bright galaxies is
consistent with the presence of significant amounts of dust mixed in with the
gas.

\section{Young Stars in the Outer Disk of M\,31}

A striking feature of the C--M diagram of Figure 8a is the
presence of a vertical sequence of blue stars with $V-I \approx 0$, well
separated from the Red Giant branch.  Previously, \citet{hod88} and
\citet{dav93} found blue stars in two different fields located at $\sim 20$ kpc
from the center of M\,31, but our field is larger and significantly further out
in the disk and extends over a range of projected galactocentric distances from
about 23 to 33 kpc.  In Figure 15 we show the distribution of
these stars plotted over the \HI\ contours.  There is a good correspondence of
the number density of these stars with the \HI\ column density.  There are even
some clusters visible.  The few remaining blue objects e.g.\ in the extreme
south-west corner are probably Galactic white dwarfs or quasars, which will
cause a small contamination over the whole image of M\,31.

A study of the C--M diagram of stars in a part of our field was carried out
earlier by \citep{ric90}.  However, their work refers to a region of low \HI\
density located near the middle of our image on the East side, where few blue
stars are to be found.

We find it somewhat surprising to have discovered evidence for star formation
which is both ongoing and wide spread in the outer disk of M\,31.  According to
the current view of the star formation process, molecular gas is therefore also
expected to be present.

A discussion of the local conditions in this part of M\,31 (gas disk surface
density and velocity dispersion) in the context of stability thresholds and
rates of star formation in galaxy disks \citep{ken89,elm95} is hampered by the
high inclination of M\,31.  The observed \HI\ column density in this region is
N(\HI) $ \approx 10^{21}$ \HI\ atoms cm$^{-2}$, only about a factor of 4 below
the column densities observed in the ``star-forming ring'' of M\,31 at about 10
kpc galactocentric radius \citep{eme74}.  Face-on values will be a factor of
$\approx 4.4$ smaller (for an inclination of $77\degr$) if a uniform slab model
is assumed for the gas disk.  The representative column densities in our outer
disk field of M\,31 are therefore about $ 2.3 \times 10^{20}$ \HI\ atoms
cm$^{-2}$.  Owing to the high inclination, the determination of intrinsic gas
velocity dispersion also suffers from averaging of gas cloud profiles along the
line of sight.  The calculation of a critical density is therefore uncertain,
but it seems likely that the observed gas surface density is well below the
threshold for forming very massive, ionizing stars.  This is in agreement with
our observations, since the young stars we have found do not seem to include
stars more luminous than about B0.

Deep H$\alpha$ images of three actively-star-forming late-type spiral galaxies
(NGC 628, NGC 1058, \& NGC 6946) provide new evidence for massive star formation
at galactocentric radial distances up to $\approx 2 \times {\rm R}_{25}$
\citep{fer98a,fer98b}.  Our results on M\,31 show that ongoing star formation
activity can also take place in the outer parts of earlier-type galaxies,
although the stars being formed are generally of lower mass.

\section{Conclusions}

We have presented evidence for the existence of appreciable amounts of dust well
mixed with the atomic gas in the far outer regions of M\,31, in an area at a
galactocentric radius of $\approx 116'$, or $\approx 1.2 \times $ R$_{25}$.  The
dust-to-atomic-gas ratio is $\gtrsim 1/3$ of the Galactic value in the solar
neighborhood.  These results refer not to isolated positions, but to large areas
of size $\sim 20$ kpc$^2$ or more in the outer disk of this galaxy.  Dust is
made of heavy elements; if primordial gas is expected to be lacking in heavy
elements (and therefore contains no dust), then the extended \HI\ gas in the
outer parts of M\,31 is not primordial.

High values of metallicity in the outer disk of M\,31 are not completely
unexpected.  \citet{jac86} measured an oxygen abundance $O/H = 3.2 \times
10^{-4}$ in an \HII\ region 17 kpc from the center of M\,31, and a value of $1.2
\times 10^{-4}$ in a planetary nebula at 33 kpc projected radius, but this
object has halo kinematics.  In our Galaxy, \citet{deg93} and \citet{dig94} have
detected CO emission from gas at a kinematic distance of 28 kpc.  This shows
that heavy elements are present in the Galaxy almost at the edge of the \HI\
disk (kinematic radius $\sim 30$ kpc).  \citet{fer98a} also report metallicities
of 0.10 - 0.15 solar in \HII\ regions in the outer parts of several late-type
spirals.  However, these detections refer to isolated positions where peculiar
local conditions may occur; our result on M\,31 refers to the outer disk gas in
general.

We have also found evidence for recent star formation in the outer parts of
M\,31.  These young stars have the signatures of B stars and appear both in
clusters and in isolation.  Their presence points to the existence of molecular
gas in the outer disk of M\,31.  The amount of molecular gas can not be directly
determined from our observations, and if it is very cold (T$_k \lesssim 5$ K)
and/or of modest density ($\lesssim$ 300 \Htwo\ molecules cm$^{-2}$), detecting
photon emission from it will be very difficult.  As an order-of-magnitude
estimate, we note that the addition of a column N(\Htwo) of molecular gas
equivalent to the observed column N(\HI) of atomic gas would reduce our observed
dust-to-atomic-gas ratio of $\gtrsim 1/3$ to a dust-to-\textit{total-gas} ratio
of $\gtrsim 1/9$, a value more in agreement with the low metallicities in the
outer parts of disk galaxies reported by \citet{fer98a}.

Finally, we have found a suggestion in our data that the ISM in the outer disk
of M\,31 is clumpy on length scales smaller than about 200 pc.  While this
result may be well-known for the ISM in the inner parts of galaxies, our data
provides the first indication that the gas in the faint outer disks of galaxies
is similarly clumpy.  A higher-resolution map of the \HI\ distribution in this
area with e.g.\ the VLA would provide the means to examine the correspondence of
gas, dust, and young stars in these outer regions of M\,31 in significantly more
detail.

\acknowledgements

The observations were collected at the Canada--France--Hawaii Telescope at Mauna
Kea.  We thank Gerard Luppino of the University of Hawaii for the loan of the
UH8K CCD mosaic camera which made the present observations possible, and the
Directors of the Canada-France-Hawaii Telescope and Space Telescope Science
Institute for financial support which facilitated the work on this paper.  The
blank field data were kindly provided by M.  Cr\'ez\'e and A.  Robin.  Helpful
comments on earlier drafts of the paper were made by Nino Panagia and Ken
Freeman.

\onecolumn

\clearpage

\begin{deluxetable}{ccc}
\tablewidth{0pt}	
\tablecaption{Counts and colors of galaxies in the SA\,57 blank field.}
\tablehead{
\colhead{V magnitude} & \colhead{Counts}
& \colhead{ $\langle V - I \rangle$ Colors}}
\startdata
$V < 22.1$ & 185 & 1.1 \\
$V < 22.5$ & 280 & 1.2 \\
$V < 24.75$ & 2025 & 1.0 \\
\enddata
\end{deluxetable}

\clearpage

\begin{figure}
\epsscale{0.8}		
\plotone{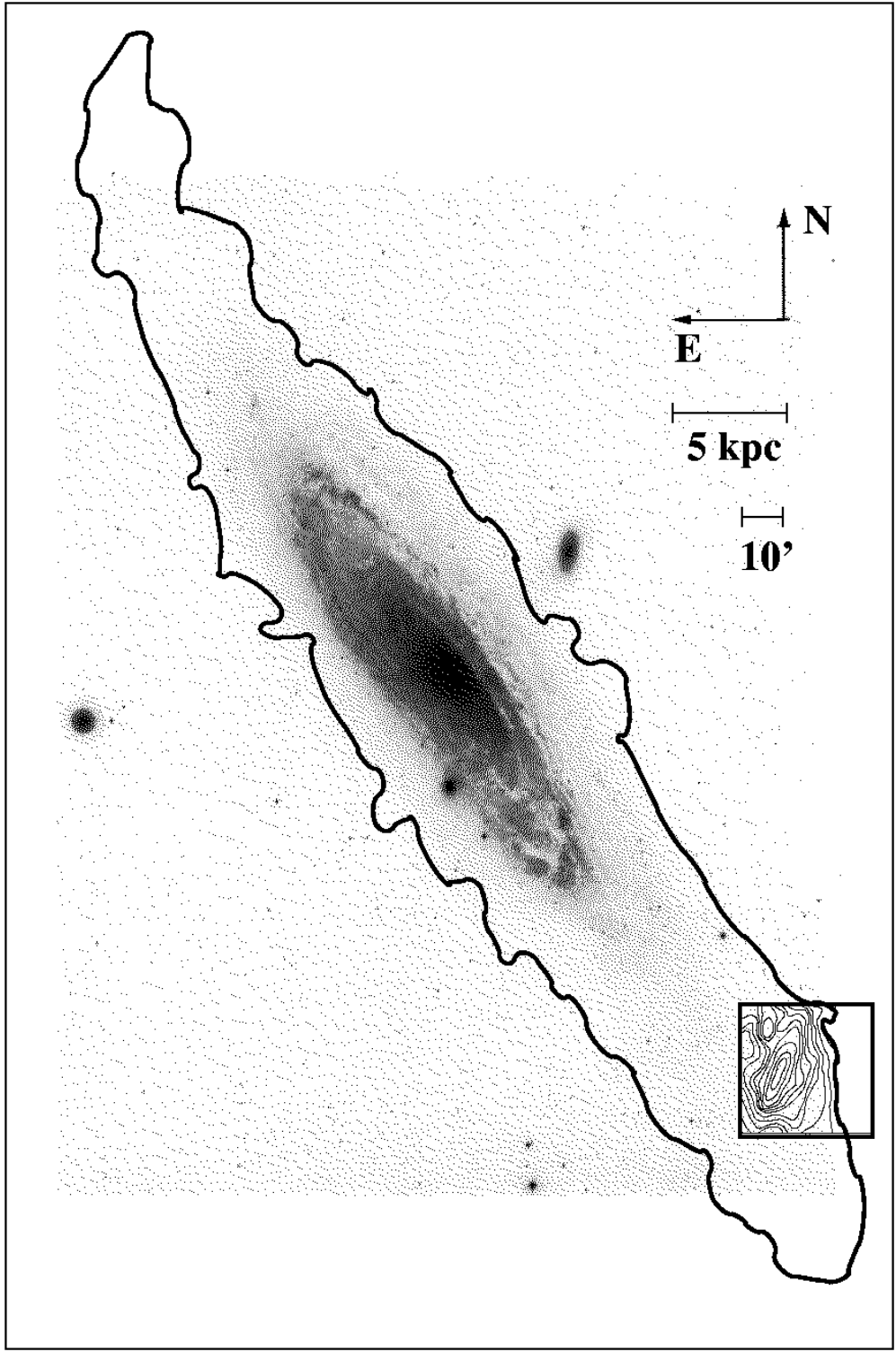}
\caption
{Sketch of the relation of the observed {\protect $28' \times 28'$} UH8K field (square box
at lower right) to the optical image of M\,31 (Sandage \& Tammann 1981).  Also shown is the
large-scale {\protect \HI\ } emission extending over {\protect $5^{\circ}$}, (thick contour, at {\protect $
\approx 8 \times 10^{19}$ atoms cm$^{-2}$}), and the higher-resolution {\protect \HI\ } map
(thin contours within the UH8K field, see Figure 11 for contour
values), from Emerson (1974) and Newton \& Emerson (1977).}
\end{figure}

\clearpage

\begin{figure}
\plotone{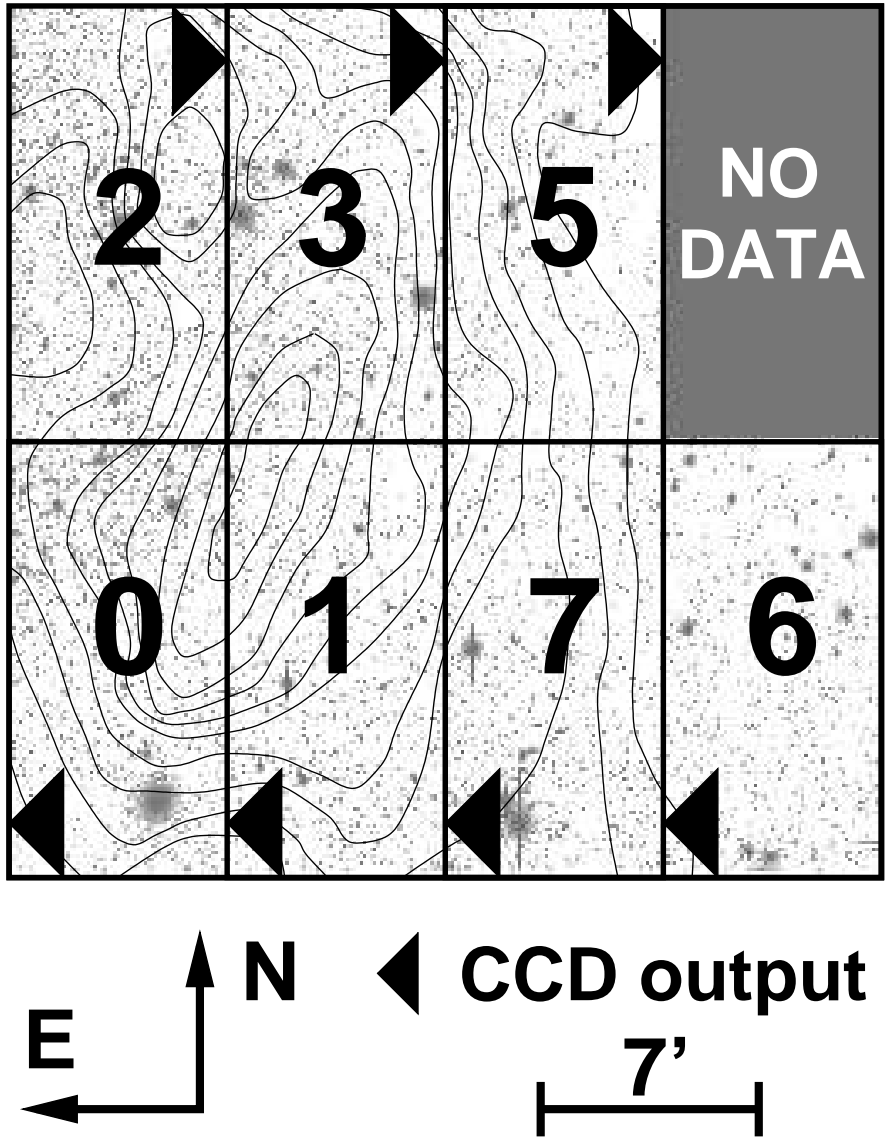}
\caption
{Sketch of the full UH8K field showing the location of the 8 CCD
chips which make up the mosaic. CCD4 has been ignored in this paper.
The arrowheads show the readout directions for each chip. Note that there
are very few bright Galactic stars contaminating this field.}
\end{figure}

\clearpage

\begin{figure}
\epsscale{0.8}		
\plotone{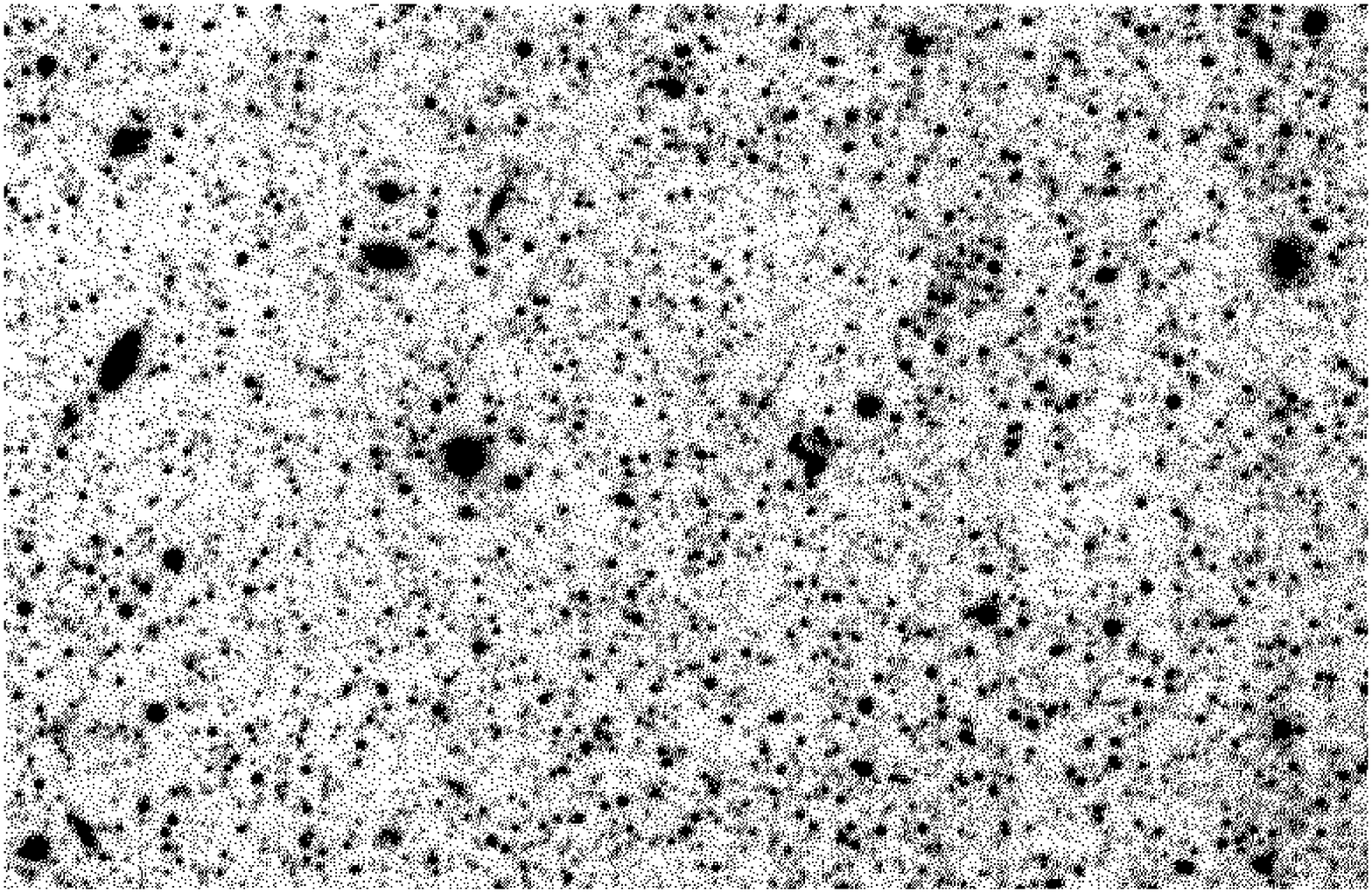}
\caption
{A portion of the final averaged V image in CCD3. This sub-image is
{\protect $700 \times 500$ pixels = $2.40' \times 1.72'$} in size and is oriented on
this page such that north is up and east is to the left.}
\end{figure}

\begin{figure}
\epsscale{0.8}		
\plotone{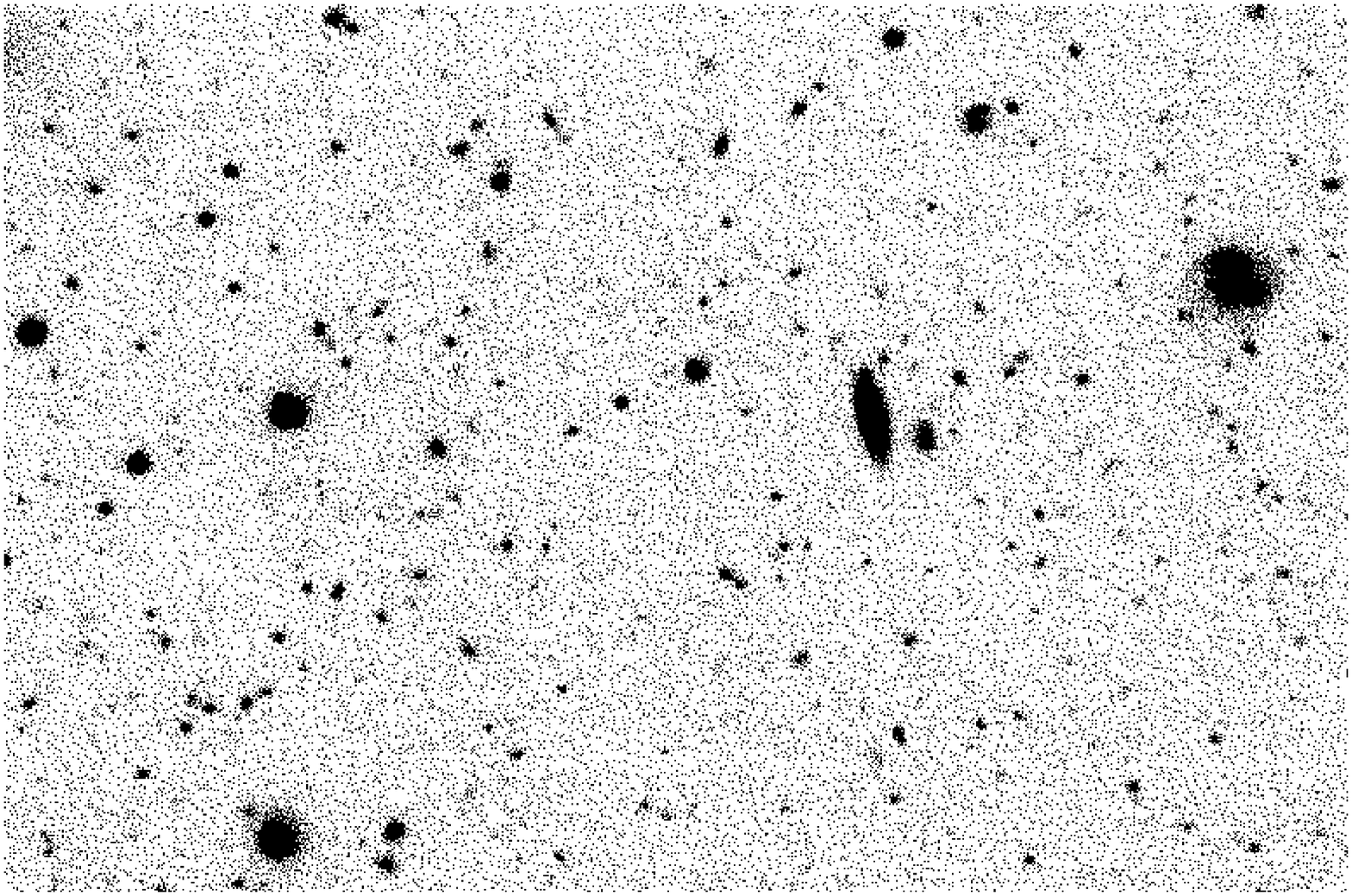}
\caption
{A portion of the final averaged V image in SA\,57. This sub-image is
{\protect $700 \times 500$ pixels = $2.40' \times 1.72'$} in size and is oriented on
this page such that north is up and east is to the left.}
\end{figure}

\clearpage

\begin{figure}
\epsscale{1.1}		
\plottwo{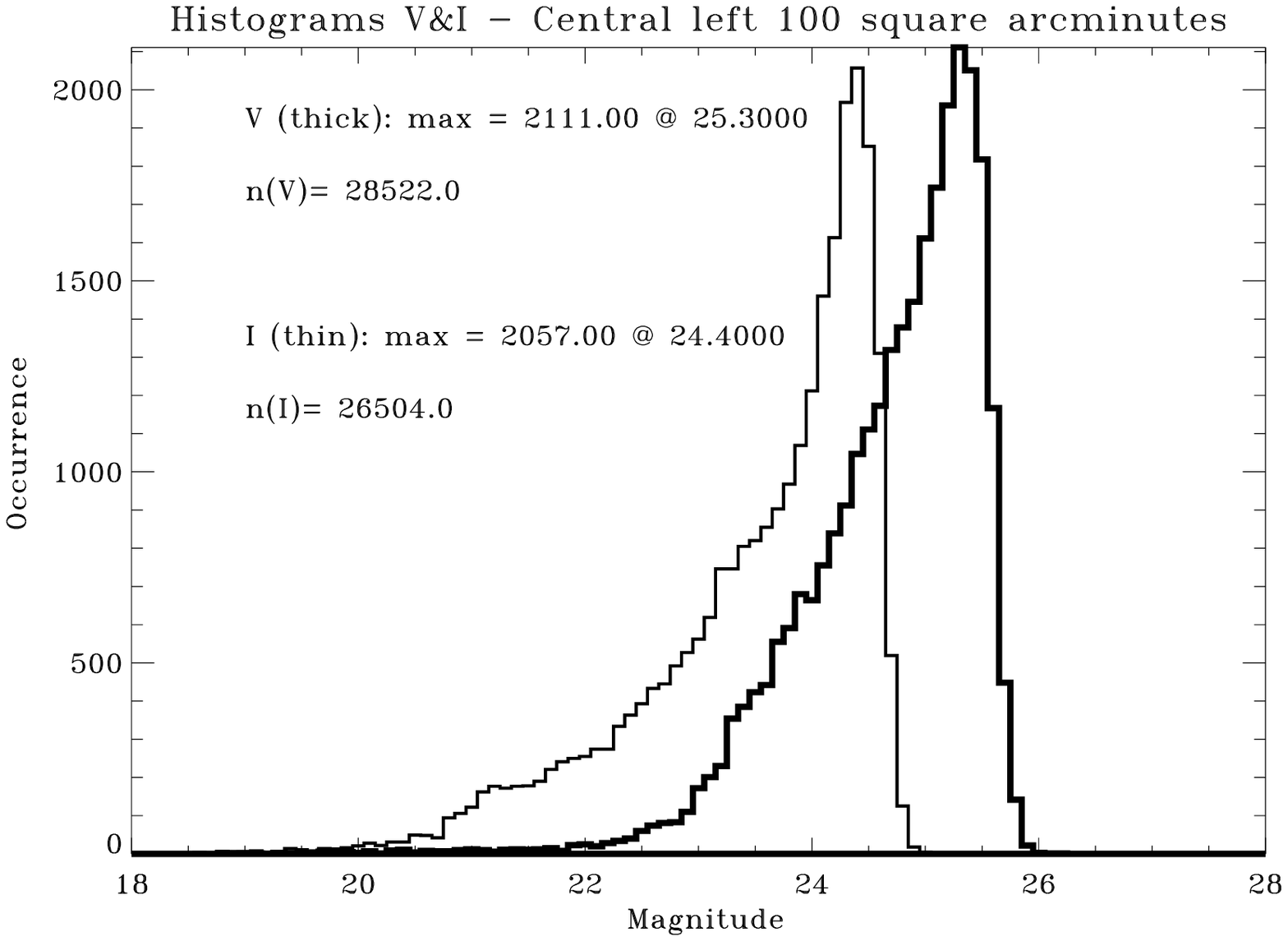}{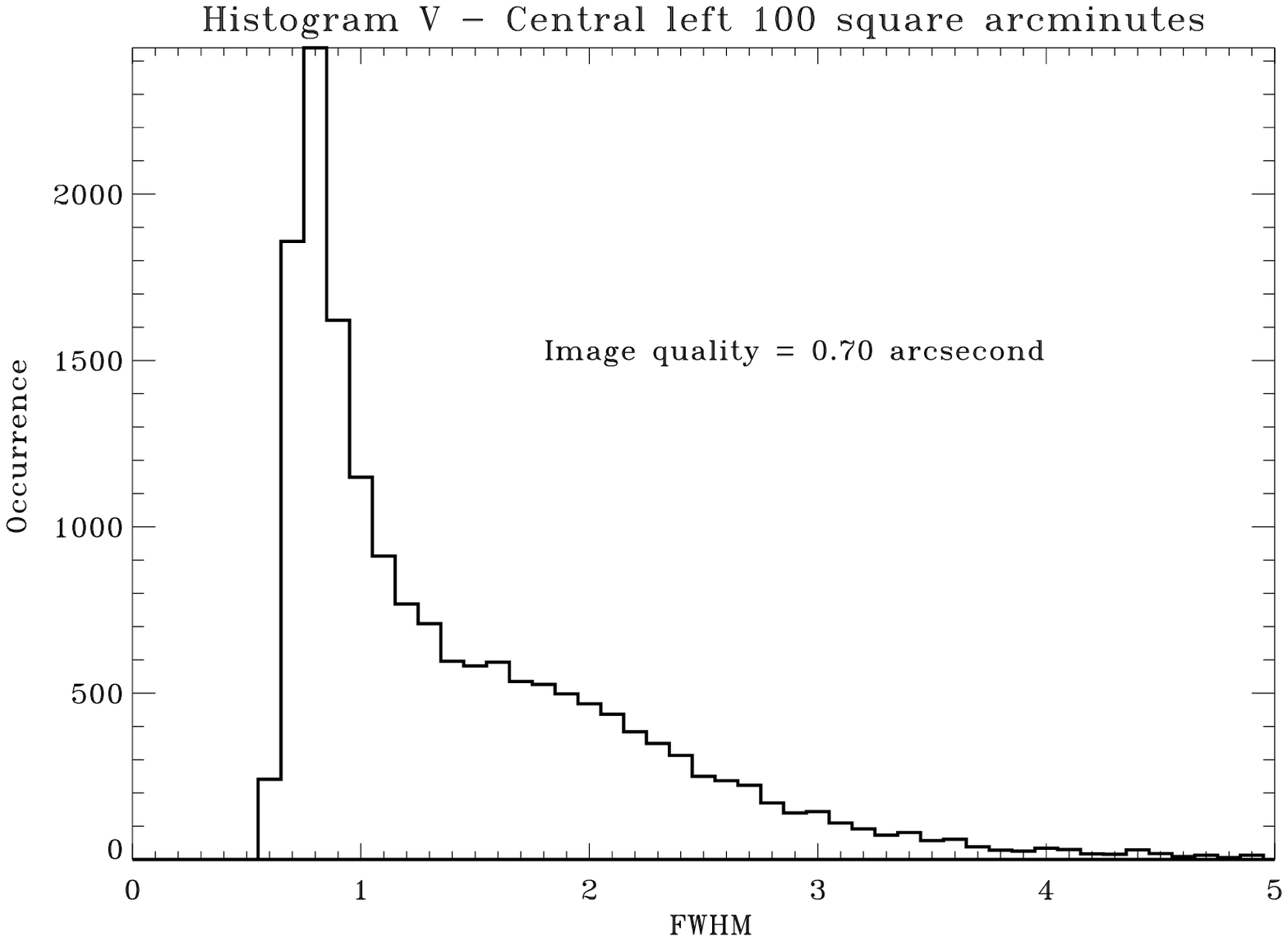}
\caption
{The histograms for all objects on a sub-field of 100 {\protect arcmin$^2$}
covering one of the CCD chips. a) (left panel) Histograms of V and I.
The smooth rise and rapid drop illustrate the high degree of
completeness of our sample, to {\protect $V < 25.4$, $I < 24.4$}. b) (right panel)
Histogram of object sizes (FWHM).}
\end{figure}


\begin{figure}
\epsscale{0.75}		
\plotone{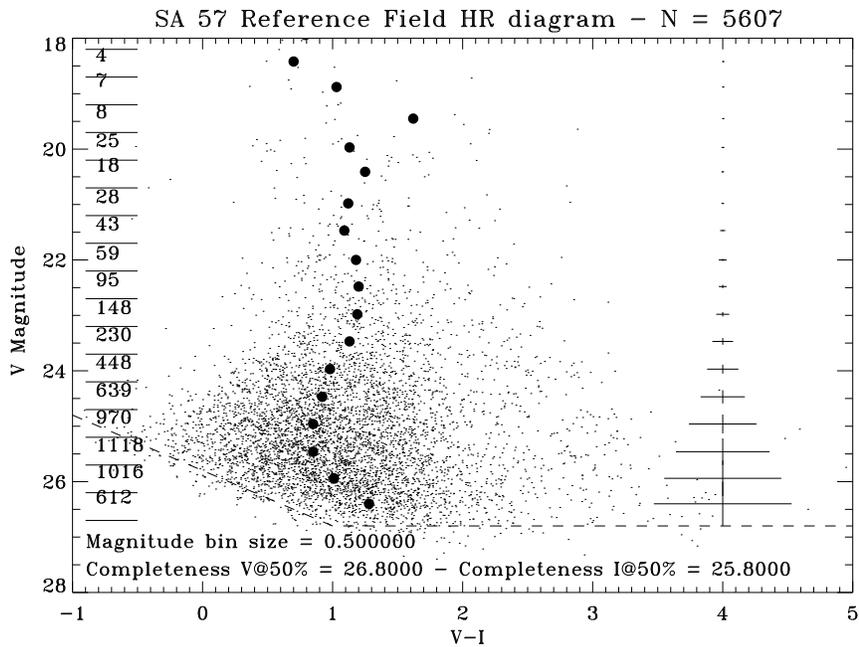}
\caption
{Color--Magnitude diagram for galaxies (objects with FWHM {\protect $> 1.5''$})
in CCD3 on the ``blank'' field SA\,57 near the Galactic North Pole.
The area covered on the sky is {\protect $7.03' \times 14.06'$}.}
\end{figure}

\clearpage

\begin{figure}
\plotone{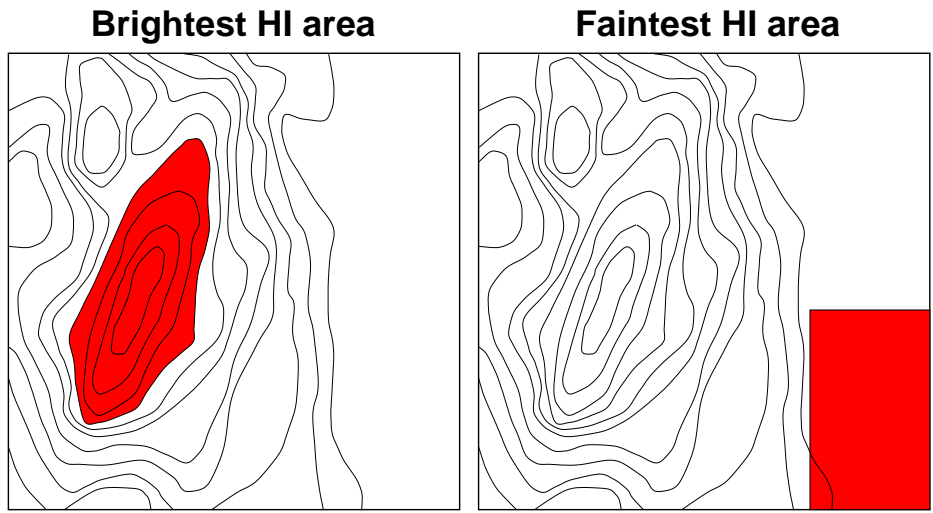}
\caption{The areas defined on the field in regions which are bright in 
{\protect \HI\ }
(left panel) and faint in {\protect \HI\ } (right panel).  The areas are the same, 70 arcmin{\protect $^2$}.  Contour levels are given in Figure 11; the
{\protect \HI}-bright area in the left panel includes contours at and above
N({\protect\HI}) = {\protect $ 1.1 \times 10^{21}$} {\protect \HI\ } atoms 
cm{\protect $^{-2}$}.}
\end{figure}

\clearpage

\begin{figure}
\plotone{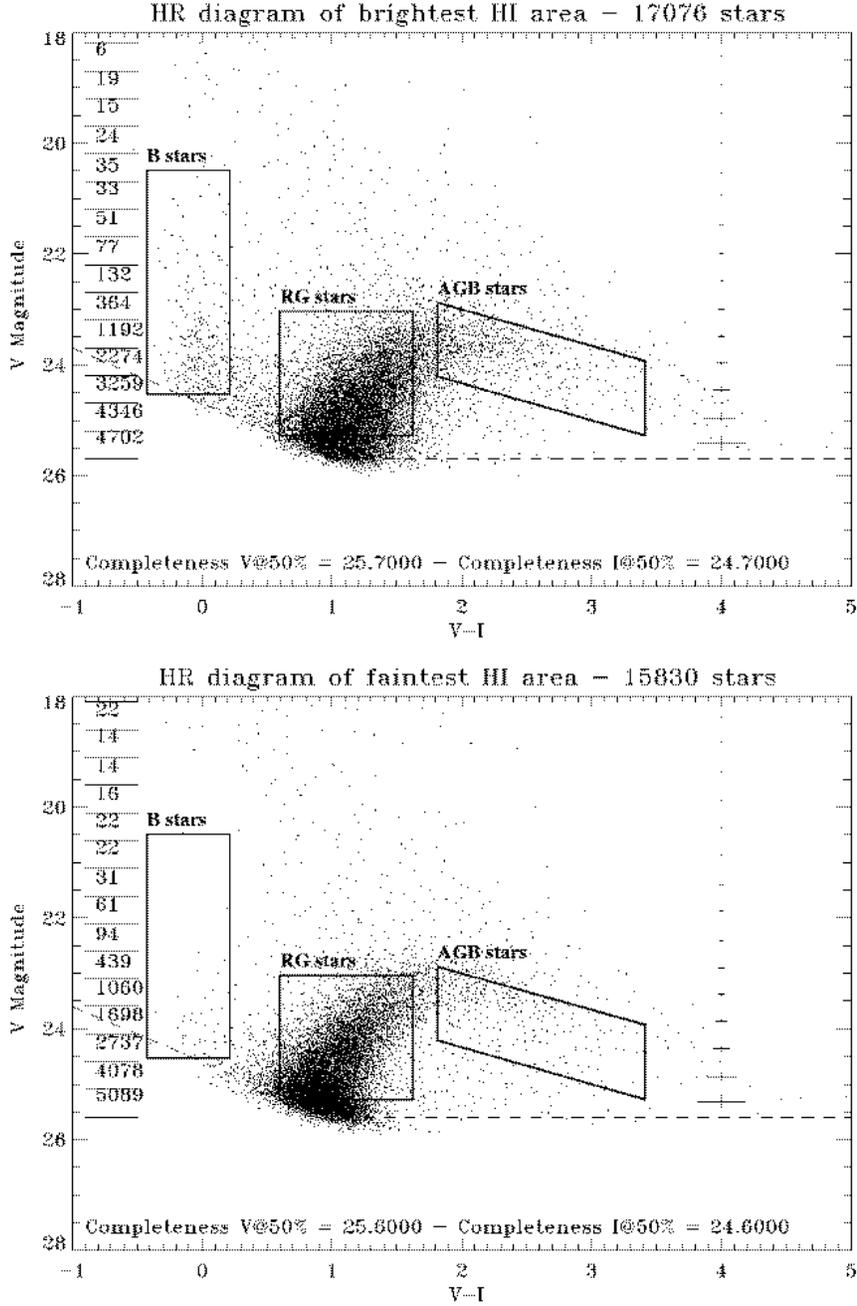}
\caption
{The {\protect V vs.\ V-I} color-magnitude diagrams for objects classified as
stars:  a) (upper panel) in the brightest {\protect \HI\ } areas of the image, as
shown in the inset; and, b) (lower panel) in the south--west part of
our field, well away from the bright {\protect \HI\ } contours. In both cases the
sub-fields are 70 {\protect arcmin$^2$} in area, and contain nearly the same
number of stars ({\protect $\approx 16,000$}).  The boxes in these figures
delineate areas of the C--M diagram populated by stars in M\,31 of
different types; B-stars,  Red Giants (RG), and Asymptotic Giant Branch
(AGB) stars. Note the shift of the data points downwards and to the
right in panel (a) compared to panel (b), a consequence of extinction
and reddening by dust mixed in with the {\protect \HI }.}
\end{figure}

\clearpage

\begin{figure}
\plotone{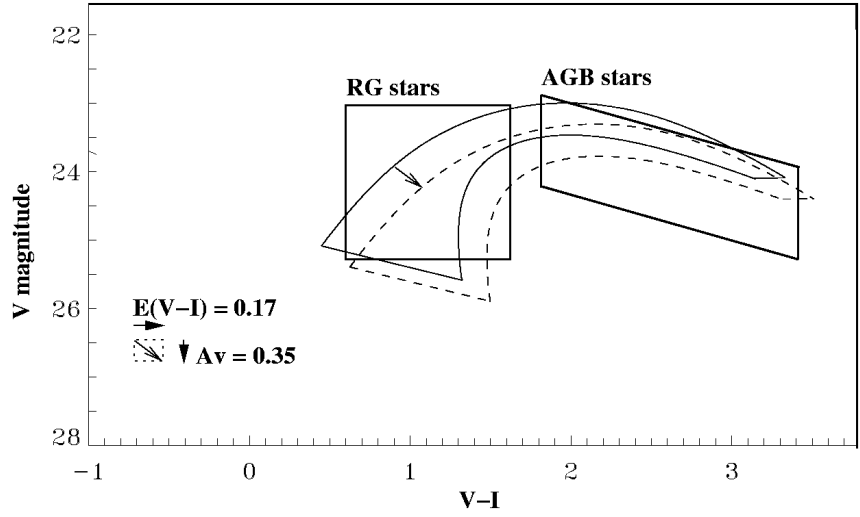}
\caption
{Sketch showing the shift of the data from Figure 8b
to Figure 8a as a result of extinction and reddening.
Our rough derivation of the values of the shifts in the two coordinates
is shown on the Figure.}
\end{figure}

\clearpage

\begin{figure}
\plotone{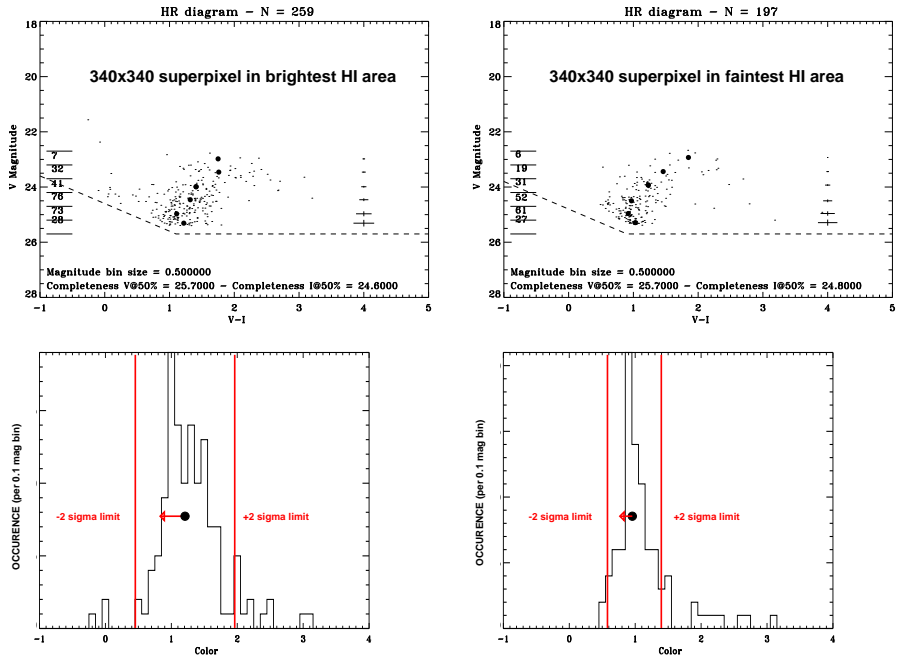}
\caption
{C--M diagrams for two representative superpixels, one in an {\protect \HI-rich } area
(upper left) and one in an {\protect \HI-poor} area (upper right). The lower panels are
the corresponding histograms from which we compute the mean V-I colors,
indicated by the large dot in the middle of the figure. The horizontal arrow
anchored to this dot shows the {\protect $1 \sigma$} dispersion.}
\end{figure}

\clearpage

\begin{figure}
\plotone{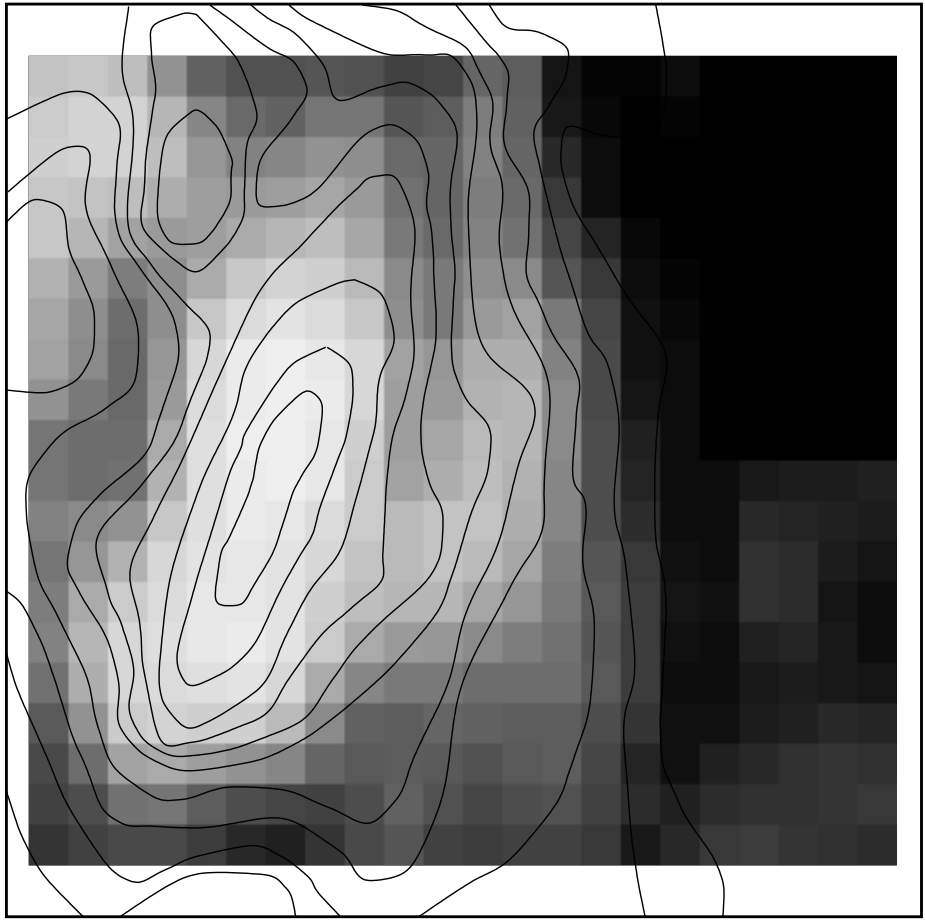}
\caption
{Map of the mean color {\protect $\langle V-I \rangle$} of stars in the field
(grey scale) compared to the column density of {\protect \HI\ } (contours).  The
field center is at {\protect $\alpha$}(1950) = 00h 33m 21.4s, {\protect $\delta$}(1950) =
39{\protect$^{\circ} 32' 21''$}. The color map is represented in ``superpixels'',
but has been smoothed to {\protect $3 \times 5$} superpixels in order to more
closely correspond with the {\protect \HI\ } resolution.
The grey scale goes from {\protect $\langle V-I \rangle = 1.05$} (dark) to
{\protect $\langle V-I \rangle = 1.26$} (white) on a linear scale.  The {\protect \HI\ } map
is from Newton \& Emerson (1977) with a resolution of {\protect $3.6' \: {\rm EW} \times
5.8' \: {\rm NS}$} FWHM, and has here been corrected for their primary
beam attenuation. The UH8K mosaic was rotated {\protect $+3.6^{\circ}$} (east
through north) to conform to the N--S orientation of the {\protect \HI\ } map.  The
{\protect \HI\ } contours are drawn at levels of 2 through 18 in steps of 2, plus
contours at 13 and 19, in units of {\protect $7.7 \times 10^{19}$} atoms
{\protect cm$^{-2}$}. Note the good correlation between reddening and the column
density of {\protect \HI}.}
\end{figure}

\clearpage

\begin{figure}
\epsscale{1.0}		
\plotone{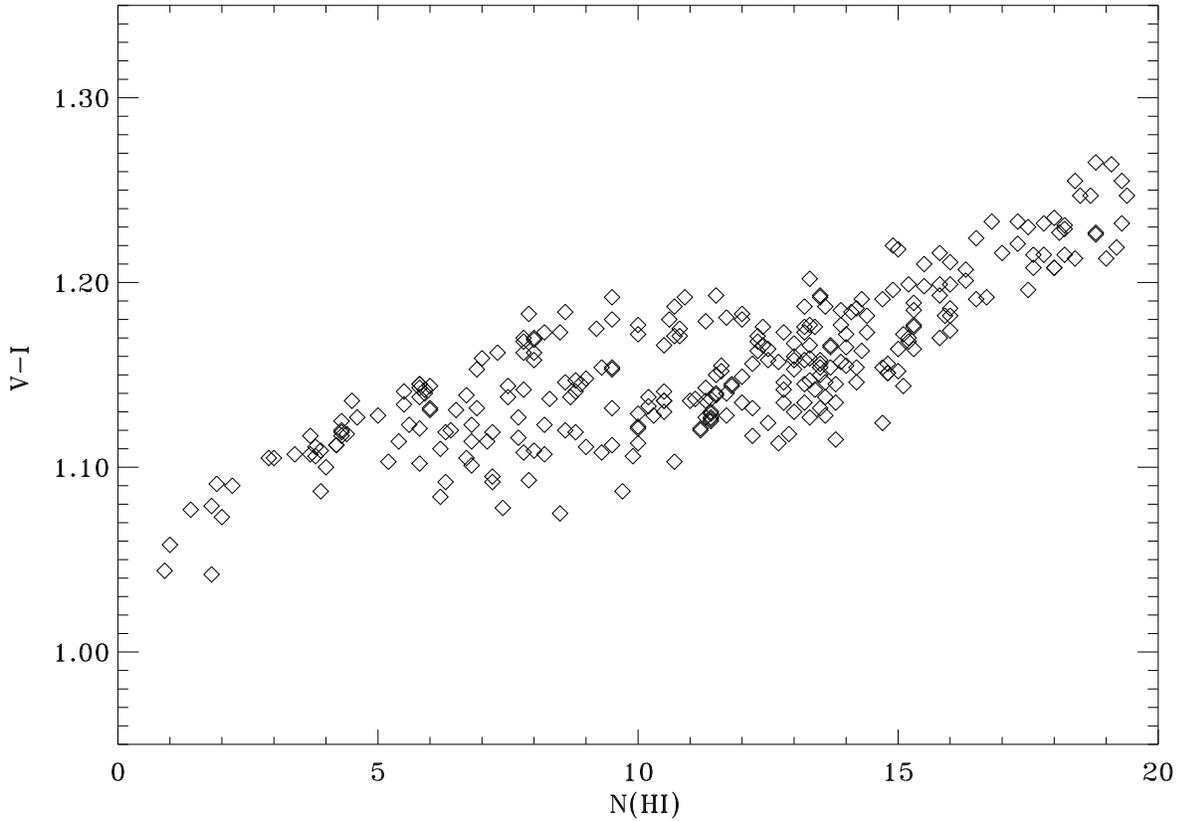}
\caption
{The relation between the mean color {\protect $\langle V-I \rangle$} (mag) of
stars and the column density of {\protect \HI\ } (in units of {\protect $7.7 \times 10^{19}$
atoms cm$^{-2}$}) corresponding to the image of Figure
11.  The 9 columns of superpixels at the right of Fig.
11 have not been used to build this correlation in order
to reduce the effects of contamination by bluer halo stars. We have
also checked that the Galactic hydrogen as depicted indirectly by the
IRAS 100 {\protect $\mu$m} map does not contaminate this diagram. Note that the
points in this figure are not independent, hence the non-gaussian
distribution.}
\end{figure}

\clearpage

\begin{figure}
\epsscale{1.0}		
\plotone{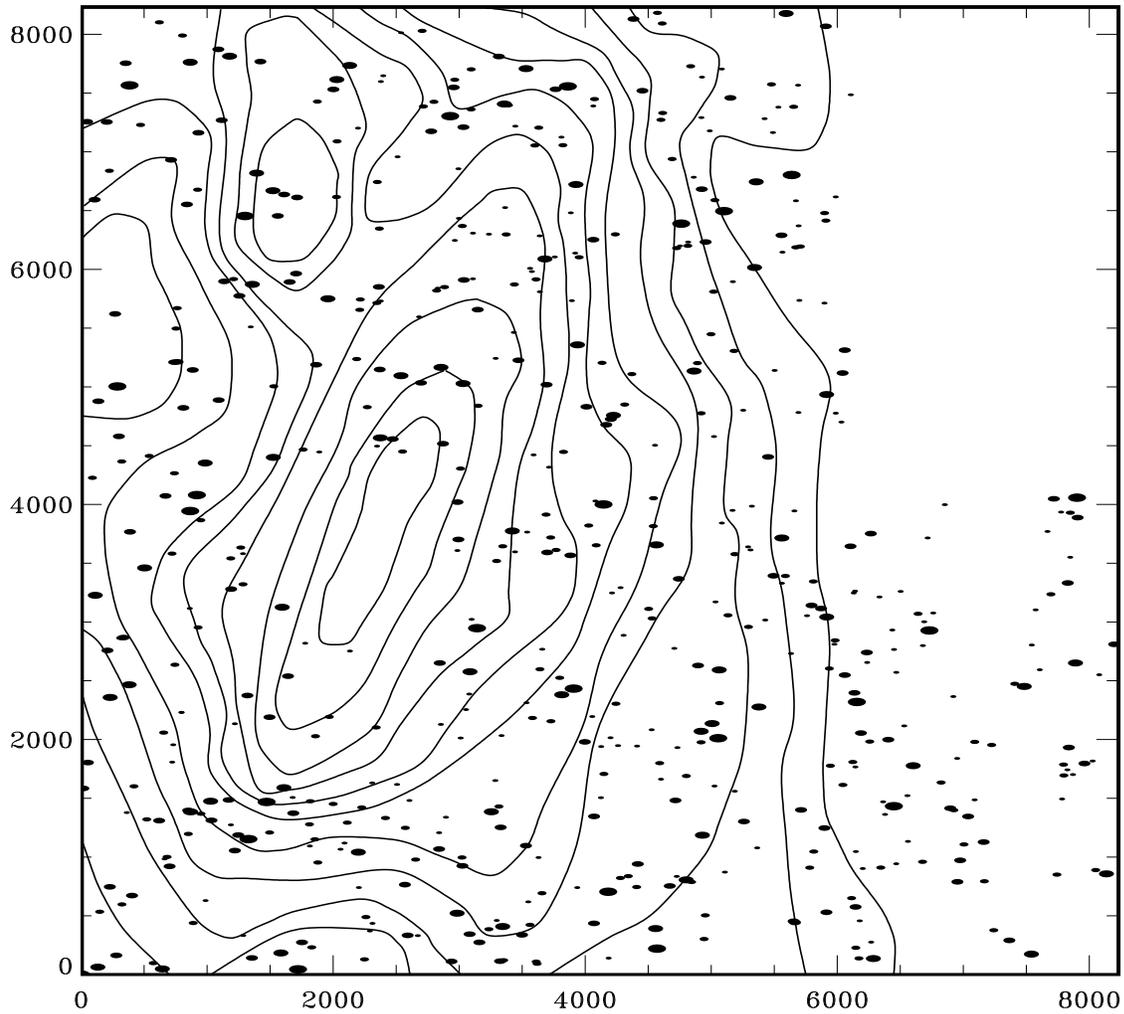}
\caption
{The distribution of ``large'' bright background galaxies over the
image as identified from a visual inspection of each CCD chip. The
largest symbol corresponds to galaxies with {\protect $V < 18.5$}. The remaining 4
smaller symbols indicate galaxy magnitudes ranging from {\protect $19 \pm 0.5$}
mag (larger symbol) to {\protect $22 \pm 0.5$} mag (smallest symbol). The contours
are N({\protect \HI}) as in Figure 11.}
\end{figure}

\clearpage

\begin{figure}
\centerline{\hbox{
\epsfig{file=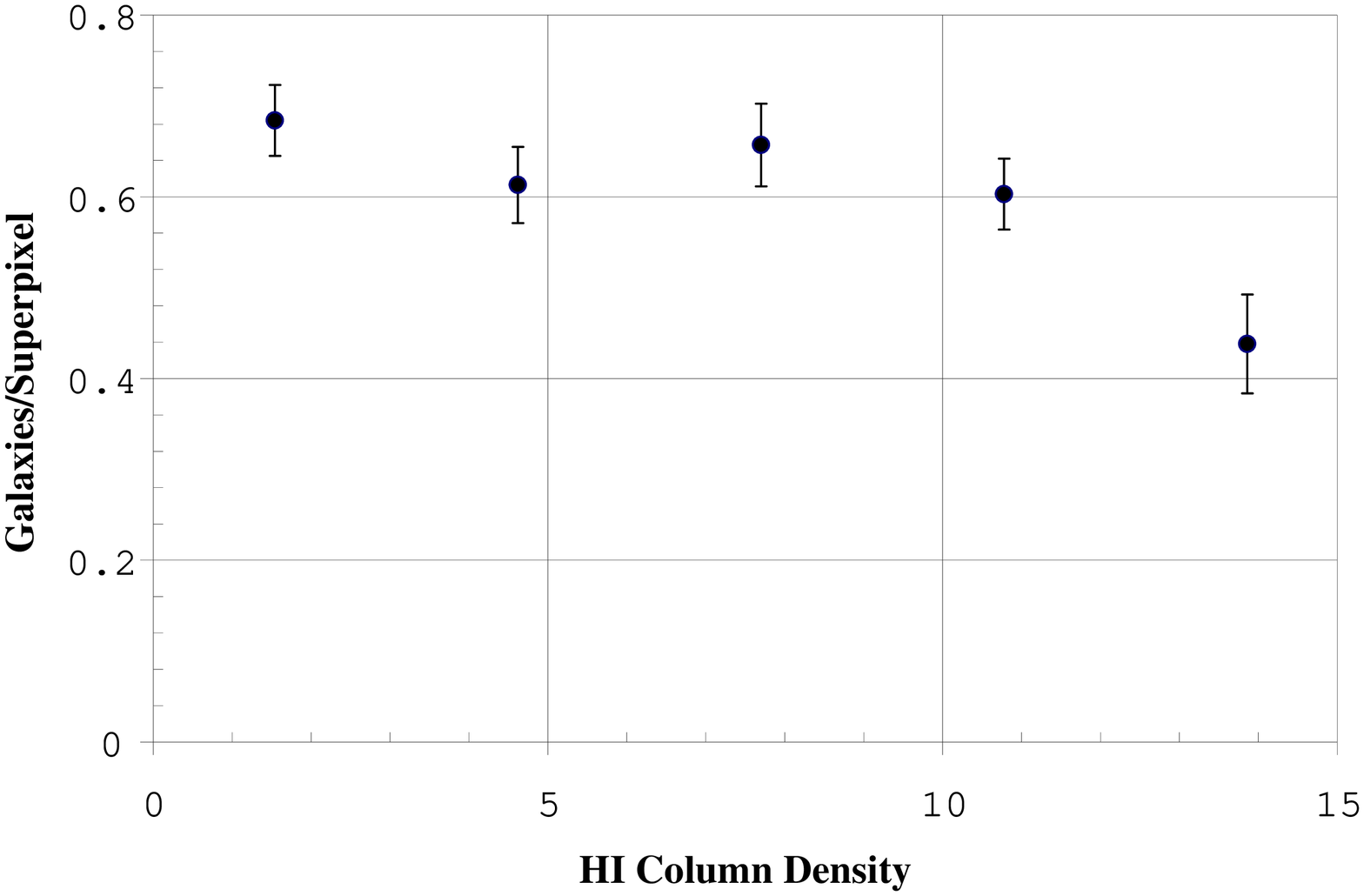,width=2.0in,angle=270}
\hspace{0.1in}
\epsfig{file=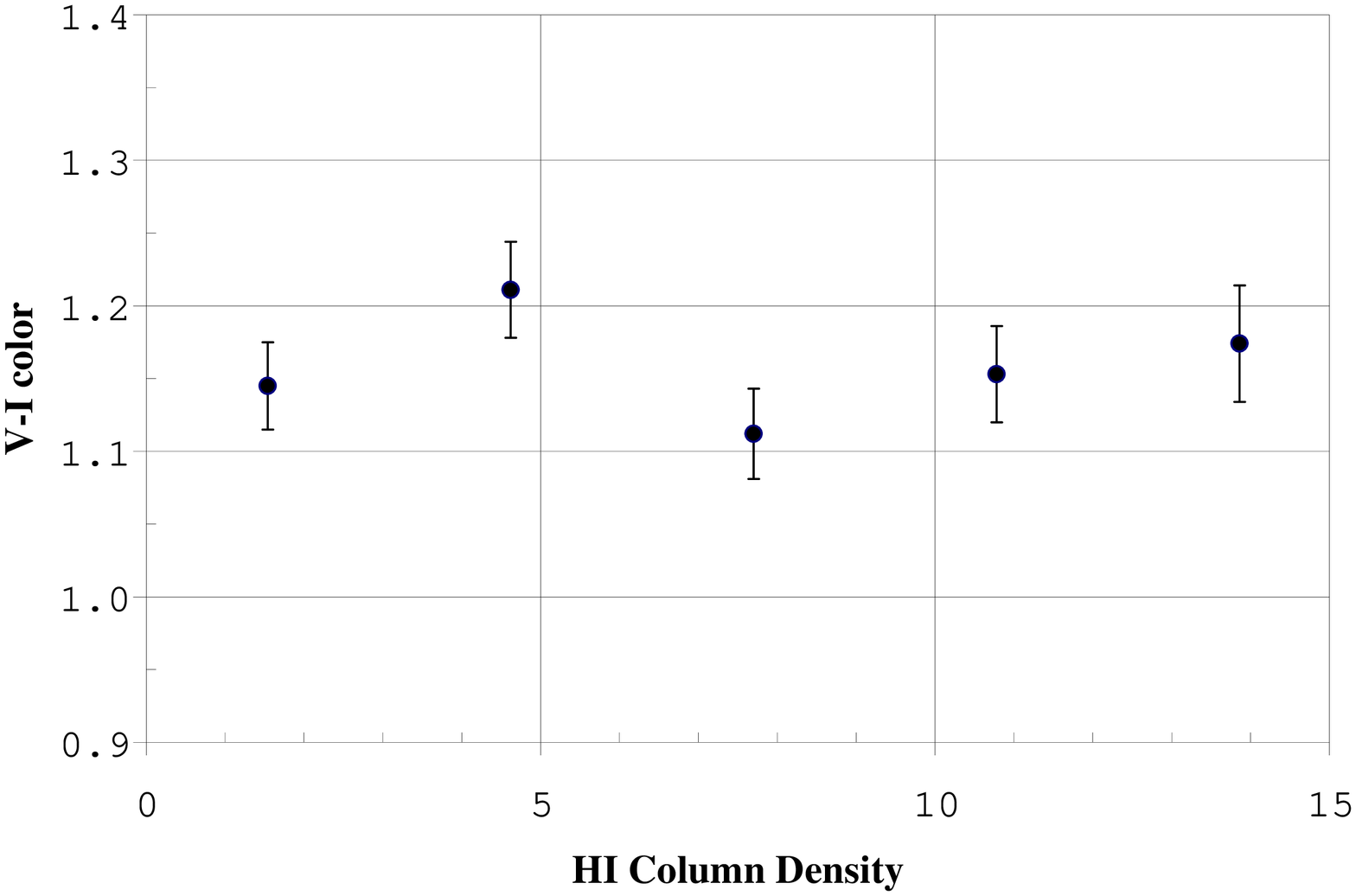,width=2.0in,angle=270}
}}
\caption
{a) (left panel) The surface density of bright background galaxies with
{\protect $V < 22.5$} in the image at different levels of atomic gas surface
density N({\protect \HI}).  The units of {\protect $N_G$} are galaxies/superpixel; a
superpixel is 1.36  arcmin{\protect $^2$}.  b) (right panel) The colors of bright
background galaxies in the image at different N({\protect \HI}) levels. The units
of N({\protect \HI}) in both panels are {\protect $1.0 \times 10^{20}$} atoms cm{\protect$^{-2}$}.}
\end{figure}

\clearpage

\begin{figure}
\plotone{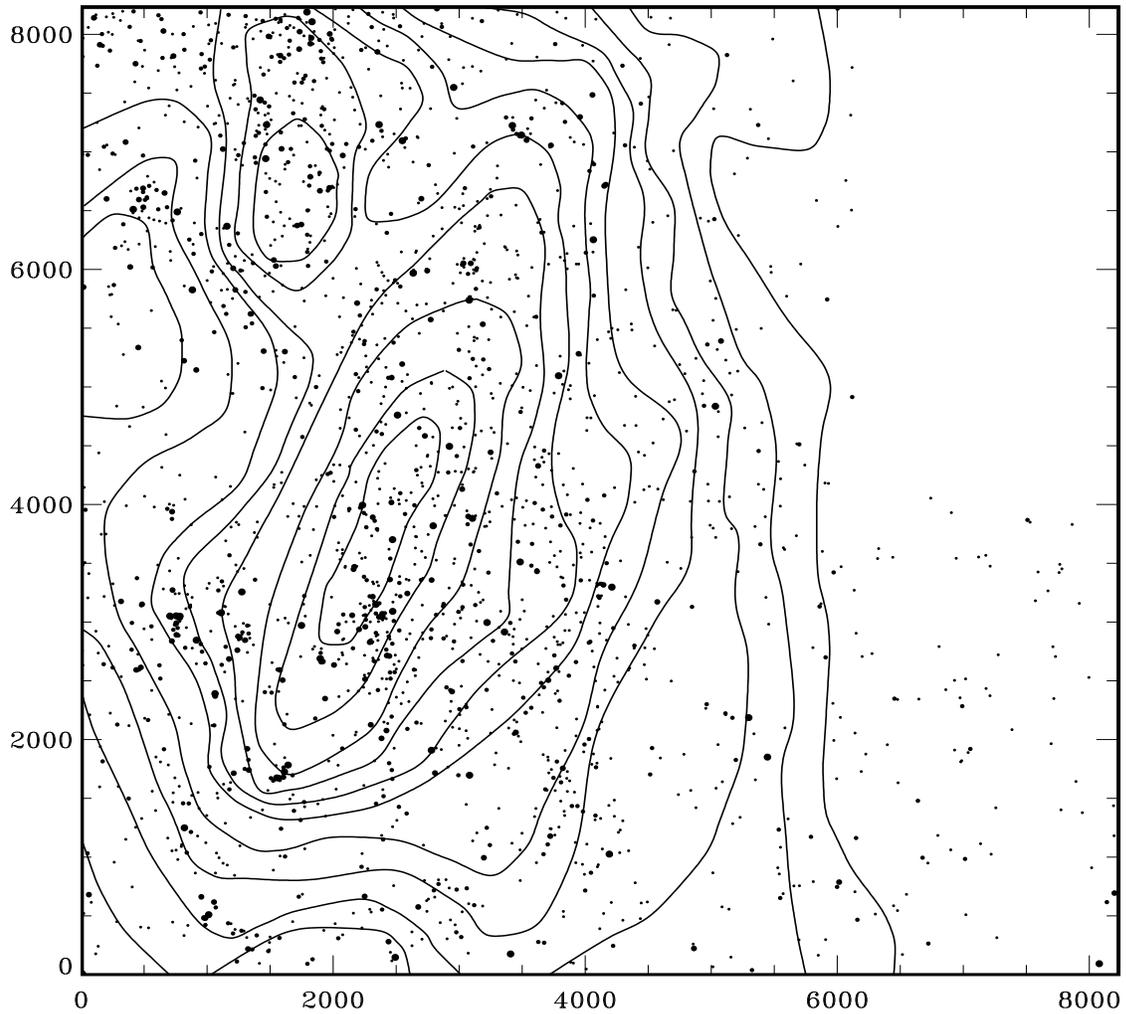}
\caption
{The distribution of the blue stars over the extreme outer disk of M\,31.  The
symbols show the locations of stars in the range {\protect $20.5 < V < 24.5$} with colors
{\protect $-0.5 < V-I < 0.2$}.  Brighter stars have larger symbols, in 4 steps of 1 mag.
The contours of {\protect \HI\ } column density are superimposed as for Figure
11.  There is some contamination by quasars and Galactic white
dwarfs, which are visible e.g.\ in the halo field at the extreme south-west of
the image.}
\end{figure}

\end{document}